\documentclass[conference]{IEEEtran}
\usepackage{epsfig,amsmath,amssymb,epsf,amsthm,scalefnt,multirow}
\usepackage{xcolor}
\usepackage{float}
\usepackage{cite}
\usepackage{psfrag}
\usepackage{graphicx}
\usepackage{subcaption}
\newsavebox{\imagebox}

\newtheorem*{lemma*}{Lemma}

\def\b0{{\pmb{0}}} 

\def\ba{{\mathbf{a}}} \def\bb{{\mathbf{b}}}  
   \def\bh{{\mathbf{h}}}

  \def\bs{{\mathbf{s}}} 
   
\def\by{{\mathbf{y}}} \def\bz{{\mathbf{z}}}

  \def\bS{{\mathbf{S}}}

\begin{document}

\title{Radar Imaging Based on IEEE 802.11ad Waveform}

 \author{\IEEEauthorblockN{Geonho Han and Junil Choi}\\
	\IEEEauthorblockA{School of Electrical Engineering\\
		Korea Advanced Institute of Science and Technology\\
		Emails: \{ghhan6,junil\}@kaist.ac.kr}}

\maketitle

\begin{abstract}
The extension to millimeter-wave (mmWave) spectrum of communication frequency band makes it easy to implement a joint radar and communication system using single hardware. In this paper, we propose radar imaging based on the IEEE 802.11ad waveform for a vehicular setting. The necessary parameters to be estimated for inverse synthetic aperture radar (ISAR) imaging are sampled version of round-trip delay, Doppler shift, and vehicular velocity. The delay is estimated using the correlation property of Golay complementary sequences embedded on the IEEE 802.11ad preamble. The Doppler shift is first obtained from least square estimation using radar return signals and refined by correcting the phase uncertainty of Doppler shift by phase rotation. The vehicular velocity is determined from the estimated Doppler shifts and an equation of motion. Finally, an ISAR image is formed with the acquired parameters. Simulation results show that it is possible to obtain recognizable ISAR image from a point scatterer model of a realistic vehicular setting.
\end{abstract}

\section{Introduction}\label{sec1}
The communication and radar systems have been developed individually so far since their performance metrics are different. The communication systems try to enable high data rates with good quality of service (QoS) while the radar systems focus on target detection and range/velocity estimation \cite{Liu:2019}. Due to the lack of available spectrum, the communication systems are soon to make use of the millimeter-wave (mmWave) spectra in practice, which may cause a coexistence problem since many radar systems, especially in vehicular environments, have already been using mmWave spectra. At the same time, the coexistence problem can be an opportunity to implement a joint radar and communication system (JRCS) using a single hardware component with reduced physical space and power consumption \cite{Mishra:2019,Choi:2016}.

The JRCS can be particularly useful for vehicle-to-everything (V2X) communications, which will become indispensable for future vehicles to ensure drivers' safety and convenience and to reduce traffic congestion through optimized driving \cite{Chen:2017,Abboud:2016}. For V2X, the radar function of JRCS would be able to provide side information to communication module for beam alignment \cite{Nuria:2016}. Although legacy vehicular sensors, e.g., radar, camera, or even light detection and ranging (LIDAR), can provide necessary information for efficient mmWave beam alignment, the JRCS with integrated radar and communication modules can be jointly optimized for vehicular environments \cite{Kumari:2017,Dokhanchi:2019}.

Recently, there have been many works on the JRCS. The typical ways to implement the JRCS are modulating communication symbols on the radar waveform or sensing targets via the communication waveform \cite{Roberton:2003,Garmatyuk:2010}. In \cite{Chiriyath:2015}, the fundamental performance bounds of data rate as a communication metric and estimation rate as a radar metric have been analyzed from an information theoretic point of view. A novel waveform has been devised for the JRCS in \cite{Kumari:2019} to improve velocity estimation accuracy with the cost of marginal data rate loss. The radar imaging exploiting JRCS with high resolution to classify the shape of object, however, has not been studied so far.

In this paper, we consider the inverse synthetic aperture radar (ISAR) imaging exploiting JRCS for a vehicular setting. The fundamental radar functions using the IEEE 802.11ad wireless local area network (WLAN) waveform have been implemented in \cite{Kumari:2017}, and we extend them to the ISAR imaging. The preamble in a single-carrier (SC) physical layer (PHY) frame of IEEE 802.11ad has an ideal characteristic suitable for target sensing \cite{Nitsche:2014}, and it can be used for ranging of targets. The least square estimation (LSE) as a straightforward approach will be exploited to recover Doppler shift. Vehicular velocity is obtained from the relation between the Doppler shifts and equation of motion for a vehicle. The proposed ISAR imaging first obtains a range profile and conducts a fast Fourier transform (FFT) after multiplying cross-range information from the estimated Doppler shifts and velocity along the cross-range direction. We verify through numerical simulations that using the proposed techniques with the commercialized IEEE 802.11ad WLAN standard can accomplish the ISAR imaging without any additional sensing hardware. Due to the extremely high carrier frequency of 60 GHz in the IEEE 802.11ad waveform, the proposed ISAR imaging is possible to achieve high resolution and precise scaling of a vehicle.


\section{System Model}\label{sec2}
\begin{figure}
	\centering
	\includegraphics[width=0.9\columnwidth]{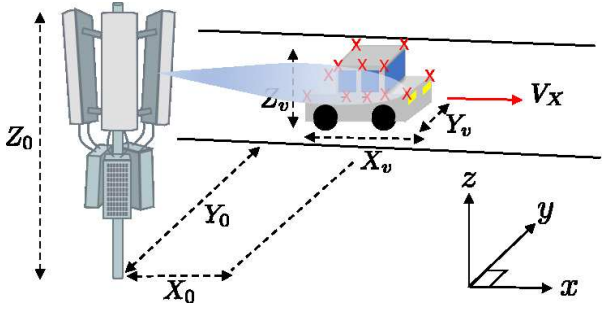}
	\caption{A vehicle-to-infrastructure (V2I) scenario considered in this paper. The distance between RSU and vehicle is assumed to be relatively larger than both sizes. The vehicle is described as its size $(X_v,Y_v,Z_v)$ and velocity $V_X$. The initial location of vehicle from the RSU is $(X_0,Y_0,Z_0)$.}\label{figure1}
\end{figure}

In this paper, the IEEE 802.11ad SC PHY frames are employed to implement the radar imaging functionality. We are interested in a situation that a roadside unit (RSU) transmits signals to a vehicle and processes the return signals at the radar module, as shown in Fig. \ref{figure1}. We start by describing a two-way multi-target radar channel model. Then, the transmit and radar received signal models of the IEEE 802.11ad waveform are developed.

\subsection{Channel Model}
We focus on the radar channel model to form the ISAR image. For the RSU to form the ISAR image of a vehicle, we further assume each vehicle consists of multiple dominant scatterers, resulting in a multi-target model.

We consider a multiple input multiple output (MIMO) system with $N_{\textrm{TX}}$ elements transmit (TX) antenna array and $N_{\textrm{RX}}$ elements receive (RX) antenna array at the RSU. A uniform planar array (UPA) is assumed to be mounted on the RSU and consists of antennas along the horizontal axis ($x$-axis) and vertical axis ($y$-axis), which means $N_{\textrm{TX}}=N_x^{\textrm{TX}}\times N_y^{\textrm{TX}}$ and $N_{\textrm{RX}}=N_x^{\textrm{RX}}\times N_y^{\textrm{RX}}$.

Assuming the TX antenna array (which works for both communication and radar functions) and the RX antenna array for the radar are co-located, the radar channel model is given as \cite{Hong:2014,Jian:2007}
\begin{align}\label{eq1}
\mathbf{H}_{\textrm{rad}}(t)&=\sum_{p=0}^{N_p-1}\sqrt{G_p(t)}\beta_p e^{j2\pi t\nu_p(t)}e^{-j2\pi f_c\tau_p(t)}\notag\\
&\qquad\qquad\times\mathbf{a}^*_{\textrm{RX}}(\phi_p(t),\theta_p(t))\mathbf{a}^{\mathrm{H}}_{\textrm{TX}}(\phi_p(t),\theta_p(t)),
\end{align}
where $\ba_{\textrm{TX}}(\phi,\theta)$ and $\ba_{\textrm{RX}}(\phi,\theta)$ represent TX and RX array steering vectors at the RSU, and $N_p$ is the number of dominant scatterers.
In (\ref{eq1}), several parameters associated with the $p$-th dominant scatterer are as follows: the large scale channel gain $G_p(t)$ that involves the effects of radar cross section (RCS) and path-loss, the small scale complex channel gain $\beta_p$ distributed according to $\mathcal{CN}(0,1)$ that is assumed to be constant during one CPI, and the azimuth and elevation angle of arrivals (AoAs) ($\phi_p(t),\theta_p(t)$). The round-trip delay is $\tau_p(t)=2r_p(t)/c$ with the distance $r_p(t)$ and the speed of light $c$. The Doppler shift of the $p$-th dominant scatterer is $\nu_p(t)=2v_p(t)/\lambda$, where $v_p(t)$ and $\lambda$ are the relative velocity and the wavelength corresponding to the carrier frequency $f_c$.

Let the spatial frequencies be
\begin{equation}\label{eq2}
\omega_x=\frac{2\pi d_x\mathrm{cos}(\theta)\mathrm{sin}(\phi)}{\lambda},\quad\omega_y=\frac{2\pi d_y\mathrm{sin}(\theta)}{\lambda},
\end{equation}
where $d_x$ and $d_y$ are the antenna spacing in the horizontal and vertical directions. We assume both antenna spacing as the half wavelength. The array steering vector is represented as
\begin{align}\label{eq3}
\ba_{\textrm{AX}}(\phi,\theta)=&\begin{bmatrix}
1, e^{j\omega_x},\cdots,e^{j(N_x^{\textrm{AX}}-1)\omega_x}\end{bmatrix}^\mathrm{T}\notag\\
&\otimes\begin{bmatrix}
1,e^{j\omega_y},\cdots,e^{j(N_y^{\textrm{AX}}-1)\omega_y}\end{bmatrix}^\mathrm{T},
\end{align}
where $\otimes$ denotes the Kronecker product, and AX $\in\left\{\textrm{TX},\textrm{RX}\right\}$.

\subsection{Transmit and Received Signal Models}
It is essential to send multiple frames to grasp the dominant scatterers with mobility. The continuous time representation of transmitted baseband signal is modeled as
\begin{equation}\label{eq4}
x(t)=\sqrt{\mathcal{E}_s}\sum_{n=-\infty}^{\infty}s[n]g_{\textrm{TX}}(t-nT_s),
\end{equation}
where $\mathcal{E}_s$ is the symbol energy, $s[n]$ is the unit energy TX symbol of IEEE 802.11ad, $g_{\textrm{TX}}(t)$ denotes the TX pulse shaping filter, and $T_s$ represents the symbol period.

To carry out the radar imaging operation, we handle only the return signals at the RX antenna array for the radar. The radar received signal that goes through a matched filter $g_{\textrm{RX}}(t)$ with the same roll-off factor for the transmitter considering the multi-target model is represented as 
\begin{equation}\label{eq5}
y(t)=\sum_{p=0}^{N_p-1}\sqrt{\mathcal{E}_s}h_p x_g (t-\tau_p(t))e^{j2\pi t\nu_p(t)}+z(t)
\end{equation}
with
\begin{equation}\label{eq6}
h_p=\sqrt{G_p(t)}\beta_p \mathbf{f}^{\mathrm{H}}_{\textrm{RX}}\mathbf{a}^*_{\textrm{RX}}(\phi_p(t),\theta_p(t))\mathbf{a}^{\mathrm{H}}_{\textrm{TX}}(\phi_p(t),\theta_p(t))\mathbf{f}_{\textrm{TX}},
\end{equation}
where $x_g(t)=\sum_{n=-\infty}^{\infty} s[n] g(t-nT_s)$ with $g(t)=g_{\textrm{TX}}(t)*g_{\textrm{RX}}(t)$ that is the linear convolution of TX and RX pulse shaping filters. The effects of noise and clutter are incorporated in $z(t)$, which is assumed to be an independent and identically distributed (IID) complex Gaussian random process with zero-mean and variance $\sigma_{\textrm{nc}}^2$. The combination of TX and RX pulse shaping filters satisfies the Nyquist criterion, i.e., $g(nT_s)=\delta[n]$. In the JRCS, $\mathbf{f}_{\textrm{TX}}$ is the TX beamformer at the communication module while $\mathbf{f}_{\textrm{RX}}$ is the RX combiner at the radar module. We assume the communication module can determine $\mathbf{f}_{\textrm{TX}}$ and $\mathbf{f}_{\textrm{RX}}$ through channel estimation and conveys $\mathbf{f}_{\textrm{RX}}$ to the radar module as in \cite{Kumari:2017}. We further assume $\mathbf{f}_{\textrm{TX}}$ and $\mathbf{f}_{\textrm{RX}}$ are fixed during one coherent processing interval (CPI). Although some parameters related to the backscattering coefficient $h_p$ in (\ref{eq6}) are varying with frames in practice, we assume $h_p$ is constant, i.e., $h_p\approx\sqrt{G_p}\beta_p \mathbf{f}^{\mathrm{H}}_{\textrm{RX}}\mathbf{a}^*_{\textrm{RX}}(\phi_p,\theta_p)\mathbf{a}^{\mathrm{H}}_{\textrm{TX}}(\phi_p,\theta_p)\mathbf{f}_{\textrm{TX}}$, during one CPI since the parameter variations are not significant to affect the radar imaging. The Doppler shift, however, is varying with frames as $\nu_p^m$.

The discrete time representation of the radar received signal in (\ref{eq5}) after sampling is given by
\begin{align}\label{eq7}
y[m,k]=\sum_{p=0}^{N_p-1}\sqrt{\mathcal{E}_s}h_p e^{j2\pi \nu_{p}^m (k+mK)T_s} s[k-\ell_p^m]+z[m,k],
\end{align}
for $m=0,1,\cdots,M-1$ and $k=\ell_0^m,\ell_0^m+1,\cdots,K-1+\ell_{N_p-1}^m$, where $m$ and $k$ denote the indices of frame and sample, and $\ell_{\alpha}^m<\ell_{\beta}^m$ for $\alpha<\beta$. Note that $M$ and $K$ are the number of frames in a CPI and the length of training samples in a frame. In (\ref{eq7}), the parameters relevant to the $m$-th frame are indicated using the superscript $(\cdot)^m$. The round-trip delay for the $p$-th dominant scatterer after sampling is expressed as
\begin{equation}\label{eq8}
\tau_p(nT_s)=mKT_s+\ell_p^mT_s+\tau^f_p(nT_s),
\end{equation}
where $\ell_p^m$ is the sampled delay that can be estimated in the received signal model while $\tau^f_p(t)$ is the unknown fractional delay after the sampling.

To build the ISAR image, we need to estimate the sampled delay $\ell_p^m$, Doppler shift $\nu_{p}^m$, and the vehicular velocity $V_X$ for the multiple dominant scatterers composing the vehicle from the return signals. The parameter estimation methods are developed in the following sections.

\section{Delay Estimation}
\begin{figure}
	\centering
	\includegraphics[width=0.9\columnwidth]{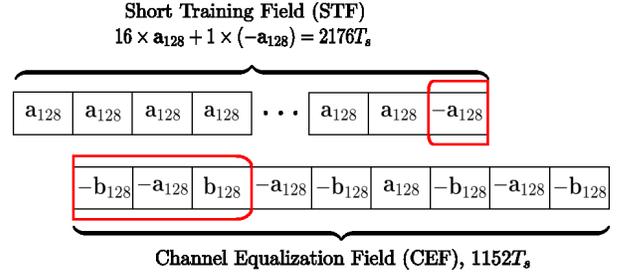}
	\caption{SC PHY preamble expressed in Golay complementary sequences. The segments in the red box are employed for delay estimation.}\label{figure2}
\end{figure}

The IEEE 802.11ad waveform employs Golay sequences, which are composed of bipolar sequences with useful correlation property, for the preamble \cite{Muns:2019}. The delay estimation comes from the intuition that the correlation property would work well to detect the dominant scatterers. We focus on several segments of the short training field (STF) and the channel equalization field (CEF) in an SC PHY frame, as shown in Fig. \ref{figure2}, where $\ba_{128}$ and $\bb_{128}$ are the Golay bipolar sequences with 128 samples. We exploit the segments in the red box since they have the perfect auto-correlation property up to $64$ and $128$ sample delays forward and backward, respectively.

Since we use the segments starting from the 2049-th sample, this offset is considered in the remaining discussions. The sampled delays of the $m$-th frame are extracted from the STF and CEF of received signals by exploiting the correlation function defined as
\begin{align}\label{eq9}
{R}_{\bs_{512}\tilde{\by}_m}[\ell]&=\sum_{k_N=0}^{511}s_{512}[k_N]\tilde{y}_m^*[\ell+k_N]\notag\\
&=\sum_{k_N=0}^{511}\sum_{p=0}^{N_p-1} \sqrt{\mathcal{E}_s}h^*_pe^{-j2\pi\nu_{p}^m(\ell+k_N+2049+mK)T_s}\notag \\
&\qquad\times s_{512}[k_N]s[k_N+2049+\ell-\ell_p^m]+\tilde{\bz}_m^{\mathrm{H}}\bs_{512},
\end{align}
where $\bs_{512}=\begin{bmatrix}\mathbf{-a}^{\mathrm{T}}_{128}\,\mathbf{-b}^{\mathrm{T}}_{128}\,\mathbf{-a}^{\mathrm{T}}_{128}\,\mathbf{b}^{\mathrm{T}}_{128}\end{bmatrix}^{\mathrm{T}}$, $\tilde{y}_m[k]=y[m,k+2049]$, and $\tilde{z}_m[k]=z[m,k+2049]$. Note that $\tilde{y}_m[k]$ and $\tilde{z}_m[k]$ are the $k$-th elements of $\tilde{\by}_m$ and $\tilde{\bz}_m$.

The radar module at the RSU can find the sampled delay from the dominant scatterer as
\begin{equation}\label{eq10}
\hat{\ell}_{\textrm{max}}^m=\textrm{argmax}_{\ell}\,\,|{R}_{\bs_{512}\tilde{\by}_m}[\ell]|.
\end{equation}
The rest of sampled delays caused by other dominant scatterers on the vehicle can be obtained by searching both forward and backward sampled delays from $\hat{\ell}_{\textrm{max}}^{m}$. The RSU can collect the sampled delays that give $|R_{\bs_{512},\tilde{\by}_m}[\ell]|$ greater than a certain threshold. From (\ref{eq9}), we set the threshold as an upper bound of the noise correlation by the Cauchy-Schwarz inequality as 
\begin{equation}\label{eq11}
\tilde{\bz}_m^{\mathrm{H}}\bs_{512}\le\|\bs_{512}\|\cdot\|\tilde{\bz}_m\|=512\cdot\sigma_{\textrm{nc}}=\sigma_{\textrm{th}}.
\end{equation}
Note that the RSU needs to search only several sampled delays from $\hat{\ell}_{\textrm{max}}^{m}$ considering the fact that the size of vehicle is less than a few meters in practice. We assume all the collected sampled delays are distinct at the first frame due to the large bandwidth of IEEE 802.11ad waveform. The collected sampled delays at the first frame are then given as
\begin{equation}\label{eq12}
\mathcal{L}_0=\left\{\hat{\ell}_0^0, \hat{\ell}_1^0,\cdots, \hat{\ell}_{\textrm{max}}^0,\cdots, \hat{\ell}_{\hat{N}_p-2}^0, \hat{\ell}_{\hat{N}_p-1}^0\right\},
\end{equation}
where $\hat{N}_p$ is the estimated number of dominant scatterers on the vehicle.

As an initial study, we assume the collected sampled delays $\mathcal{L}_m$ is frame invariant, i.e., $\mathcal{L}_m=\mathcal{L}_0$ for all $m$ of interest. This is usually the case in practice due to the short CPI; however, it is still possible that $\mathcal{L}_m$ varies with frames depending on the values of the sampled fractional delay $\tau^f_p(nT_s)$ in (\ref{eq8}), and multiple sampled delays could overlap in $\mathcal{L}_m$ as time evolves. We refer to \cite{Han:2020} how to resolve the problem of overlapped sampled delays.

\section{Doppler Shift Estimation}
Before estimating the Doppler shift $\nu_p^m$, we first need to obtain the backscattering coefficient $h_p$ defined in (\ref{eq6}). To do this, we focus on the first frame $m=0$ as
\begin{align}\label{eq13}
y[0,u]&=\sum_{p=0}^{N_p-1}\sqrt{\mathcal{E}_s}h_p e^{j2\pi \nu_p^0uT_s}s[u-\ell_p^0]+z[0,u]\notag\\
&\stackrel{(a)}{\approx}\sum_{p=0}^{\hat{N}_p-1}\sqrt{\mathcal{E}_s}s[u-\hat{\ell}_p^0]h_p+z[0,u],
\end{align}
for $u=\hat{\ell}_{0}^0,\hat{\ell}_{0}^0+1,\cdots,K-1+\hat{\ell}_{\hat{N}_p-1}^0$, where $(a)$ is from the approximation $e^{j2\pi\nu_p^0uT_s}\approx 1$, which holds due to the short symbol period $T_s$ of IEEE 802.11ad waveform. Note that $\hat{\ell}_0^0$ indicates the first return sample the radar module has received. By concatenating $K+\hat{\ell}_{\hat{N}_p-1}^0-\hat{\ell}_{0}^0$ received samples, we have
\begin{equation}\label{eq14}
\by_0=\bS\bh+\bz_0,
\end{equation}
where
\begin{align}\label{eq15}
\by_0&=\begin{bmatrix}y[0,\hat{\ell}_{0}^0],y[0,\hat{\ell}_{0}^0+1],\cdots,y[0,K-1+\hat{\ell}_{\hat{N}_p-1}^0]\end{bmatrix}^{\mathrm{T}},\notag\\
\bh&=\begin{bmatrix}\sqrt{\mathcal{E}_s}h_0,\sqrt{\mathcal{E}_s}h_1,\cdots,\sqrt{\mathcal{E}_s}h_{\hat{N}_p-1}\end{bmatrix}^{\mathrm{T}},\notag\\
\bz_0&=\begin{bmatrix}z[0,\hat{\ell}_{0}^0],z[0,\hat{\ell}_{0}^0+1],\cdots,z[0,K-1+\hat{\ell}_{\hat{N}_p-1}^0]\end{bmatrix}^{\mathrm{T}},
\end{align}
and $\bS\in\mathbb{Z}^{(K+\hat{\ell}_{\hat{N}_p-1}^0-\hat{\ell}_{0}^0)\times\hat{N}_p}
$ consists of $s[\hat{\ell}_{0}^0+\alpha-1-\hat{\ell}_{\beta-1}^0]$ as its ($\alpha,\beta$)-th element. The backscattering coefficient vector is obtained through LSE as
\begin{equation}\label{eq16}
\hat{\bh}=(\bS^\mathrm{H}\bS)^{-1}\bS^\mathrm{H}\by_0.
\end{equation}

The Doppler shift estimation can be done similarly with the backscattering coefficient estimation, but the assumption, $e^{j2\pi\nu_{p}^m (k+mK)T_s}\approx1$, may not hold when $m\neq 0$. Instead, using the fact $k\ll mK$ for a posterior frame, we fix $k=(\hat{\ell}_{0}^{m}+\hat{\ell}_{\hat{N}_p-1}^m+K-1)/2$ on the exponential term, which is the midpoint of observed samples. Then, (\ref{eq7}) is approximated~as
\begin{align}\label{eq17}
y[m,u]&\approx\notag\\
\sum_{p=0}^{\hat{N}_p-1}&\sqrt{\mathcal{E}_s}s[u-\hat{\ell}_p^0]h_p e^{j2\pi\nu_{p}^m((\hat{\ell}_{0}^m+\hat{\ell}_{\hat{N}_p-1}^m+K-1)/2+mK)T_s}\notag\\
&\qquad\qquad\qquad\qquad\qquad\qquad\qquad\,\,+z[m,u],
\end{align}
for $u=\hat{\ell}_{0}^m,\hat{\ell}_{0}^m+1,\cdots,K-1+\hat{\ell}_{\hat{N}_p-1}^m$. By concatenating all the received samples, we have
\begin{equation}\label{eq18}
\by_m=\bS\bh_m+\bz_m,
\end{equation}
where $\by_m$ and $\bz_m$ can be defined similarly as in (\ref{eq15}), and the $p$-th element of $\bh_m$ is represented as
\begin{equation}\label{eq19}
h_m[p]=\sqrt{\mathcal{E}_s}h_p e^{j2\pi\nu_{p}^m((\hat{\ell}_{0}^m+\hat{\ell}_{\hat{N}_p-1}^m+K-1)/2+mK)T_s}.
\end{equation}
The solution of LSE for linear equation in (\ref{eq18}) is given as
\begin{equation}\label{eq20}
\hat{\bh}_m=(\bS^\mathrm{H}\bS)^{-1}\bS^\mathrm{H}\by_m,
\end{equation}the estimated Doppler shift using 
and (\ref{eq16}) and (\ref{eq20}) is written~as
\begin{equation}\label{eq21}
\hat{\mathrm{\nu}}_{p}^m=\frac{\angle(\hat{h}_m[p]/\hat{h}[p])}{2\pi ((\hat{\ell}_{0}^m+\hat{\ell}_{\hat{N}_p-1}^m+K-1)/2+mK)T_s},
\end{equation}
for $p=0,1,\cdots,\hat{N}_p-1$.

The estimated Doppler shift in (\ref{eq21}) is not valid for small $m$ due to the assumption used in (\ref{eq13}). Instead, we employ Doppler shift difference to estimate the Doppler shift for small $m$, e.g., $m=0$. Here, the constant velocity of vehicle during one CPI lead to fixed Doppler shift difference between the two consecutive frames. To compute the difference, we use two properly separated frames as
\begin{equation}\label{eq22}
\boldsymbol{\mathrm{\Delta}}_{\nu}^m=\frac{\hat{\boldsymbol{\mathrm{\nu}}}^m-\hat{\boldsymbol{\mathrm{\nu}}}^{m-i}}{i},
\end{equation}
where $\hat{\boldsymbol{\nu}}^m$ is the $\hat{N}_p\times 1$ vector containing the Doppler shifts for all dominant scatterers in the $m$-th frame. In (\ref{eq22}), the frame difference $i$ needs to be carefully selected. With too small $i$, it may cause insufficient reduction of estimation error, which comes from the LSE of (\ref{eq16}) and (\ref{eq20}), in (\ref{eq21}). With large $i$, on the contrary, the estimation error can be decreased significantly; however, it may induce different number of wrappings on the phases for $\hat{\nu}_p^m$ and $\hat{\nu}_p^{m-i}$, especially when the vehicle is moving fast and experiencing high Doppler shift since the range of phase is $[-\pi,\pi]$. This phase uncertainty must be resolved before computing (\ref{eq22}).

To resolve the phase uncertainty, we first define the uncertainty corrector as
\begin{equation}\label{eq23}
c_p=|\hat{\nu}_{p}^m|-|\hat{\nu}_{p}^{m-i}|,
\end{equation}
for $p=0,1,\cdots,\hat{N}_p-1$. We need to identify how many times the wrappings happen in the Doppler shift. The corrector for the true phases becomes
\begin{align}\label{eq24}
\check{c}_p&=|\psi_p^m D_m|-|\psi_p^{m-i} D_{m-i}|\notag\\
&=|(2\pi M_p+\psi_{p,\textrm{wrap}}^{m})D_m|-|(2\pi M_p+\psi_{p,\textrm{wrap}}^{m-i})D_{m-i}|,
\end{align}
where $\psi_p^m$ is the true phase (in radian) of the $p$-th dominant scatterer at the $m$-th frame, $D_m$ denotes the inverse of denominator in (\ref{eq21}), $\psi_{p,\textrm{wrap}}^{m}$ is the remaining phase after the wrapping, and $M_p$ is the integer representing the number of wrappings happened in the Doppler shift. From (\ref{eq24}), it is easy to show that
\begin{align}\label{eq25}
M_p=\begin{cases}\frac{c_p}{2\pi (D_{m-i}-D_m)}+\frac{\check{c}_p}{2\pi (D_{m-i}-D_m)},&\textrm{for}\quad\textrm{Case\,1}\\
-\frac{c_p}{2\pi (D_{m-i}-D_m)}-\frac{\check{c}_p}{2\pi (D_{m-i}-D_m)},&\textrm{for}\quad\textrm{Case\,2}\\
\frac{c_p}{2\pi (D_{m-i}-D_m)}-\frac{\check{c}_p}{2\pi (D_{m-i}-D_m)},&\textrm{for}\quad\textrm{Case\,3}\\
-\frac{c_p}{2\pi (D_{m-i}-D_m)}+\frac{\check{c}_p}{2\pi (D_{m-i}-D_m)},&\textrm{for}\quad\textrm{Case\,4}
\end{cases}
\end{align}
where the four cases are $(\psi_p<0,\,\psi_{p,\textrm{wrap}}>0)$, $(\psi_p>0,\,\psi_{p,\textrm{wrap}}<0)$, $(\psi_p>0,\,\psi_{p,\textrm{wrap}}>0)$, and $(\psi_p<0,\,\psi_{p,\textrm{wrap}}<0)$ in order.
Since it is not possible to know $\check{c}_p$ in practice, we use an approximated $M_p$ as
\begin{align}\label{eq26}
\bar{M}_p\approx\begin{cases}\lfloor\frac{c_p}{2\pi (D_{m-i}-D_m)}\rceil,&\textrm{for}\quad\psi_{p,\textrm{wrap}}>0\\
\lfloor-\frac{c_p}{2\pi (D_{m-i}-D_m)}\rceil,&\textrm{for}\quad\psi_{p,\textrm{wrap}}<0
\end{cases}
\end{align}
where $\lfloor\cdot\rceil$ maps the argument into its nearest integer. The phase wrapping effect is compensated using $\bar{M}_p$ as
\begin{equation}\label{eq27}
\hat{\hat{\nu}}_{p}^m=\hat{\nu}_{p}^m+2\pi\bar{M}_p D_m,\quad \hat{\hat{\nu}}_{p}^{m-i}=\hat{\nu}_{p}^{m-i}+2\pi\bar{M}_p D_{m-i}.
\end{equation}
The Doppler shift difference $\boldsymbol{\Delta}_{\nu}^m$ in (\ref{eq22}) is then computed using $\hat{\hat{\nu}}_p^m$ and $\hat{\hat{\nu}}_p^{m-i}$. The median of $\boldsymbol{\Delta}_{\nu}^m$, $\Delta_{\nu,\textrm{med}}$, is used as all dominant scatterers' Doppler shift difference as
\begin{align}\label{eq28}
\hat{\hat{\boldsymbol{\nu}}}^u&=\hat{\hat{\boldsymbol{\nu}}}^m+(u-m)\Delta_{\nu,\textrm{med}}\mathcal{\mathbf{1}},
\end{align}
for $u=0,1,\cdots,M-1$, where $\mathcal{\mathbf{1}}$ denotes the all-ones vector. Although arbitrary $m$ might work for computing the Doppler shift difference, we set $m=M-1$ to minimize the midpoint approximation error in (\ref{eq17}).

\begin{figure}
	\centering
	\includegraphics[width=0.7\columnwidth]{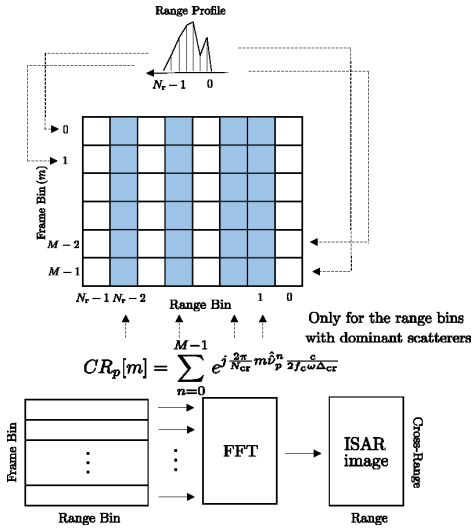}
	\caption{ISAR image formation procedure where $CR_p[m]$ contains the information for the cross-range positions.}\label{figure3}
\end{figure}

\section{ISAR Image Formation}
Before creating the ISAR image, the range reconstruction should be preceded. This is simply done by assigning the estimated backscattering coefficient $\hat{h}[p]$ to the range bin corresponding to the estimated sampled delay of $p$-th dominant scatterer. The estimated Doppler shift $\hat{\hat{\nu}}_p^m$ is transformed into the cross-range information by multiplying $c/(2f_c\omega\Delta_{\mathrm{cr}})$, where $\omega$ denotes the rotational velocity, the cross-range resolution is $\Delta_{\mathrm{cr}}=\lambda W_D/(2M\omega)$ with the Doppler frequency bandwidth $W_D=2\omega Y_{\textrm{size}}f_c/c$ and the size of image projection plane $X_{\textrm{size}}\times Y_{\textrm{size}}$ \cite{Ozdemir:2012}. To show the ISAR image that indicates cross-range positions of the dominant scatterers during $M$ frames, $CR_p[m]$, defined in Fig. \ref{figure3}, is multiplied with the backscattering coefficient $\hat{h}[p]$, which is placed only for the range bins with the dominant scatterers in advance. Finally, the ISAR imaging is accomplished via FFT along the cross-range direction. The overall procedure of ISAR image formation is represented in Fig. \ref{figure3}, where the numbers of range and cross-range bins are $N_{\mathrm{r}}=\lfloor{X_{\textrm{size}}/\Delta_{\mathrm{r}}\rfloor}$ and $N_{\mathrm{cr}}=\lfloor{Y_{\textrm{size}}/\Delta_{\mathrm{cr}}\rfloor}=M$ with $\lfloor{\cdot\rfloor}$ that rounds its argument to the lower nearest integer, the range resolution $\Delta_{\mathrm{r}}=c/(2W)$, and the bandwidth $W$.

\begin{figure}
	\centering
	\includegraphics[width=0.5\columnwidth]{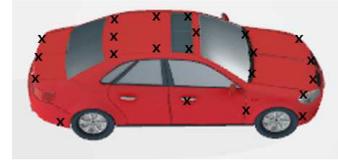}
	\caption{Point scatterer model of a realistic vehicle (side view), where the number of dominant scatterers is $N_p=22$.}\label{figure4}
\end{figure}

\begin{figure}
	\centering
	\includegraphics[width=0.63\columnwidth]{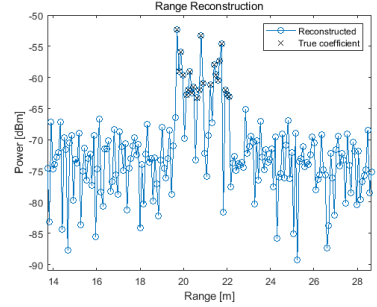}
	\caption{Range reconstruction using estimated delays and backscattering coefficients.}\label{figure5}
\end{figure}

To transform the estimated Doppler shift $\hat{\hat{\nu}}_p^m$ into cross-range information, the RSU needs to know the vehicular velocity $V_X$ for $\omega$, where $\omega=V_X^{\perp}/R_0$ with the velocity component perpendicular to the range direction $V_X^{\perp}=V_X\mathrm{cos}(\mathrm{tan}^{-1}(X_0/Y_0))$ and the distance between the RSU and vehicle $R_0=\sqrt{X_0^2+Y_0^2+Z_0^2}$. The velocity $V_X$ is obtained using the definition of Doppler shift and an equation of motion as
\begin{align}\label{eq29,30}
\nu_{p}^m&=\frac{2V_X \mathrm{sin}(\phi_p^m)}{\lambda}\stackrel{(b)}{\approx}\frac{2V_X \phi_p^m}{\lambda},\\
\textrm{CPI}\cdot V_X&=R_0 (\mathrm{sin}(\phi_p^0)-\mathrm{cos}(\phi_p^0)\mathrm{tan}(\phi_p^{M-1}))\notag\\
&\stackrel{(b)}{\approx}R_0(\phi_p^0-\phi_p^{M-1}),
\end{align}
where any $p$ is possible since all the dominant scatterers on the vehicle experience the same velocity $V_X$. The small angle approximation in $(b)$ is from the long distance between the RSU and vehicle. The azimuth angles in (30) are substituted with the estimated Doppler shifts using (29). Accordingly, the velocity is approximately estimated as
\begin{equation}\label{eq32}
\hat{V}_X\approx\sqrt{\frac{\lambda R_0(\hat{\hat{\nu}}_{p}^0-\hat{\hat{\nu}}_{p}^{M-1})}{2\cdot\textrm{CPI}}}.
\end{equation}

\section{Simulation Results}
In this section, we perform simulations to verify the effectiveness of proposed ISAR imaging via the IEEE 802.11ad waveform. The TX (for both communication and radar) and radar RX antenna arrays at the RSU are UPAs consisting of $8\times8$ elements. The IEEE 802.11ad specification states $f_c=$ 60 GHz carrier frequency, $W=$ 1.76 GHz bandwidth, 3328 training samples in a frame, root-raised cosine (RRC) filter with roll-off factor 0.25 for pulse shaping filters, and $T_f=13632T_s$ time interval of one frame with $T_s=1/W$ \cite{Spec:2012}. The initial location of vehicle from the RSU is $(X_0, Y_0, Z_0) =$ (0 m, 20 m, -7 m). The size and velocity of vehicle are $(X_v, Y_v, Z_v) =$ (5 m, 2 m, 1.5 m) and $V_X=$ 40 m/s. Other simulation parameters are defined as follows: the RCS of the vehicle $\sigma_{\textrm{RCS}}=$ 20 dBsm, the linear scale RCS per dominant scatterer $10^{(\sigma_{\textrm{RCS}}/10)}/N_p$ m$^2$, $\textrm{CPI} = 10$ ms, the path-loss exponent 2, the image projection plane size $X_{\textrm{size}}\times Y_{\textrm{size}}=$ 15 m $\times$ 20 m, and the number of frames during one CPI $M=\lfloor \textrm{CPI}/T_f\rfloor$. The transmitted signal power at the RSU is 30 dBm. The noise spectral density $N_o$ and average clutter power $P_c$ are assumed -174 dBm/Hz and -72.275 dBm, which result in $\sigma_{\textrm{nc}}^2=N_oW+P_c$. We set $i=6$, which determines the gap of two frames for the Doppler shift difference in (\ref{eq22}).

\begin{figure}
	\begin{subfigure}{0.46\linewidth}
		\centering
		\includegraphics[width=\linewidth]{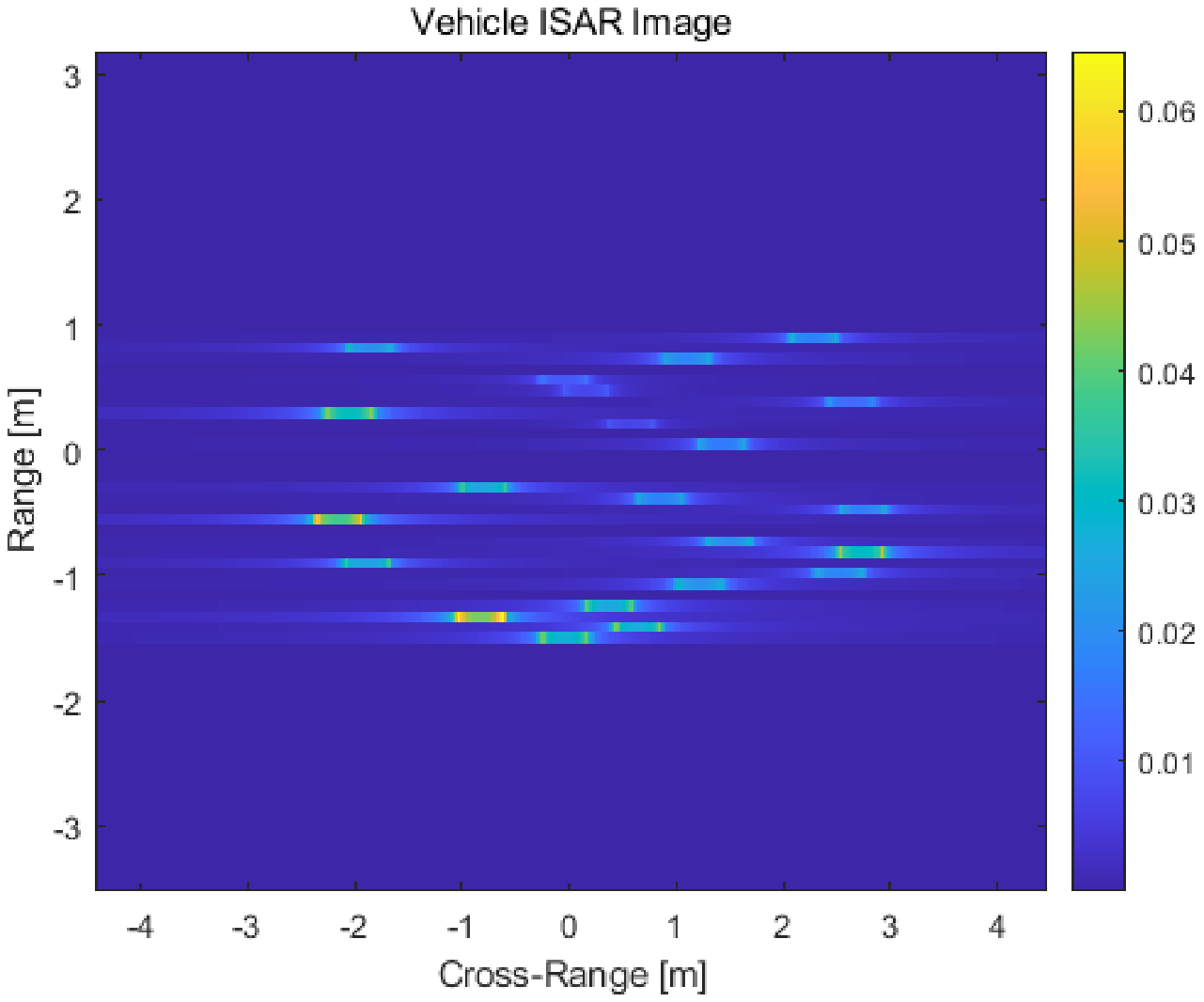}
		\caption{}
		\label{figure6(a)}
	\end{subfigure}\hfill
	\begin{subfigure}{0.46\linewidth}
		\centering
		\includegraphics[width=\linewidth]{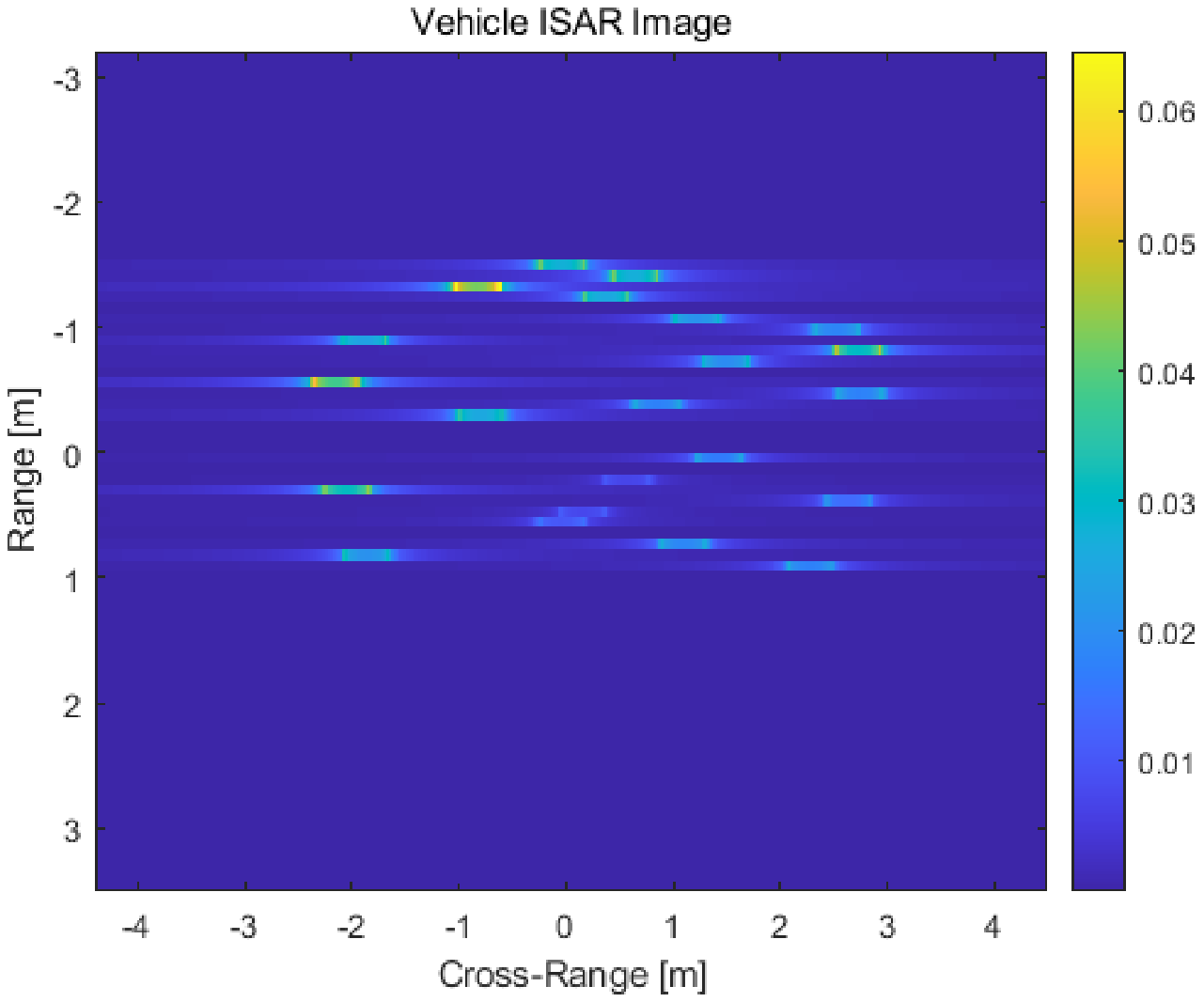}
		\caption{}
		\label{figure6(b)}
	\end{subfigure}
	\caption{The ISAR image (a) and flipped version (b).}\label{figure6}
\end{figure}

We assume the vehicle as a point scatterer model as shown in Fig. \ref{figure4}. First, we show the range reconstruction result at $m=0$ in Fig. \ref{figure5} using the estimated sampled delays $\mathcal{L}_0$ and the backscattering coefficients $\hat{\bh}$. The figure clearly shows that almost perfect range reconstruction is possible. The range resolution is about 8.52 cm due to the broadband of IEEE 802.11ad, which is enough to distinguish most of multiple dominant scatterers in the range profile.

In Fig. \ref{figure6(a)}, we present the ISAR image of vehicle. The image projection plane is a surface including the rotational velocity direction and the direction that the RSU looks at the moving vehicle, which makes the ISAR image flipped compared to the original point scatterer model. By flipping the original image as in Fig. \ref{figure6(b)}, the RSU can obtain the vehicular image that is very close to the model in Fig. \ref{figure4}.

\section{Conclusion}
In this paper, we proposed the ISAR imaging technique based on the JRCS for vehicular environments using the commercialized IEEE 802.11ad waveform. The range profile of multiple dominant scatterers on a vehicle was accurately reconstructed using the extremely high carrier frequency and the good correlation property of preamble. The Doppler shift was then estimated by means of the LSE based on the assumption of constant velocity of the vehicle during sufficiently short CPI. After obtaining the velocity using the equation of motion for the vehicle, the RSU was able to form the ISAR image exploiting the estimated parameters. We demonstrated the effectiveness of the proposed ISAR imaging technique through simulations with realistic model of vehicular environments.

\section*{Acknowledgment}
This work was supported by Institute of Information $\&$ communications Technology Planning $\&$ Evaluation (IITP) grant funded by the Korea government (MSIT) (No.2020-0-01882, The Development of Dangerous status recognition Platform in building based on WiFi Low Power Wireless signal sensing without sensor or video camera) and by the National Research Foundation of Korea (NRF) grant funded by the MSIT of the Korea government (No. 2019R1C1C1003638).

\bibliographystyle{IEEEtran}
\bibliography{refs_all}

\end{document}